\documentclass[a4paper,journal]{IEEEtran} 
\IEEEoverridecommandlockouts
\usepackage[dvipsnames]{xcolor}
\usepackage{cite}
\usepackage{lipsum}
\usepackage{amsmath,amssymb,amsfonts, bm, url}
\usepackage{graphicx}
\usepackage{textcomp}
\usepackage{xcolor}
\usepackage{float}
\usepackage{cite}
\usepackage{placeins}
\usepackage{textcomp}
\usepackage{multirow}
\usepackage{xcolor}
\usepackage{wrapfig}
\usepackage{algpseudocode}
\usepackage{algorithm}
\usepackage{caption}
\usepackage{subcaption}
\usepackage{comment}
\usepackage[noabbrev]{cleveref}

\usepackage{tabularx, multirow,multicol,colortbl,array}
\usepackage{graphicx,color, array}
\usepackage[dvipsnames]{xcolor}
\usepackage{float}
\usepackage{comment,psfrag}
\usepackage{enumitem}
\usepackage{tikz,tikzscale,bigints,pgfplots}
\pgfplotsset{compat=1.18}
\usetikzlibrary{positioning, arrows.meta, colorbrewer, calc, fit, matrix}
\usepackage{multirow}

\usepackage{import}
\usepackage{pgfplots}
\usetikzlibrary{calc}
\usetikzlibrary{spy}
\usetikzlibrary{matrix}
\usetikzlibrary{backgrounds}
\pgfplotsset{compat = newest}

\usetikzlibrary{shapes.misc, positioning}
\usetikzlibrary{shadows,arrows,shapes,calc,positioning,decorations.pathreplacing,fit}
\def\BibTeX{{\rm B\kern-.05em{\sc i\kern-.025em b}\kern-.08em
		T\kern-.1667em\lower.7ex\hbox{E}\kern-.125emX}}

\makeatletter
\def\endthebibliography{%
  \def\@noitemerr{\@latex@warning{Empty `thebibliography' environment}}%
  \endlist
}
\makeatother

\begin{document}

\pgfplotsset{
    standard/.style={
    axis line style = thick,
    grid = both,
    }
}

\pagestyle{empty}

\title{Blind Turbo Demodulation for Differentially Encoded OFDM with 2D Trellis Decomposition

}

\author{
    \IEEEauthorblockN{ Chin-Hung Chen$^{\star}$}
    , \IEEEauthorblockN{Yan Wu$^{\ddagger}$}%
    , \IEEEauthorblockN{Wim van Houtum$^{\star\ddagger}$}%
    , and \IEEEauthorblockN{Alex Alvarado$^{\star}$}%
    \\
    \IEEEauthorblockA{\textit{$^{\star}$Information and Communication Theory Lab, Eindhoven University of Technology, The Netherlands}}\\
    \IEEEauthorblockA{\textit{$^{\ddagger}$NXP Semiconductors, Eindhoven, The Netherlands}}\\
    \IEEEauthorblockA{c.h.chen@tue.nl}
}

\maketitle
\thispagestyle{empty}

\begin{abstract}
    Digital Audio Broadcasting (DAB)-like systems employ differentially encoded (DE) phase-shift keying (PSK) for transmission. While turbo-DE-PSK receivers offer substantial performance gains through iterative decoding by making the DE-PSK an inner code, they rely on accurate channel estimation without pilots, which is a key challenge in DAB-like scenarios. This paper develops a fully blind turbo-DE-PSK scheme that jointly estimates channel phase, channel gain, and noise variance directly from the received signal. The design leverages a two-dimensional (2D) trellis decomposition for blind phase estimation, complemented by power-based estimators for channel gain and noise variance. We provide a comprehensive system assessment across practical system parameters, including inner code length, phase quantization, and 2D block size. Simulation results show that the blind 2D turbo demodulator approaches the performance of receivers with perfect channel knowledge and remains robust under realistic transmission conditions.


\end{abstract}


\section{Introduction} \label{sec:intro}
Orthogonal frequency-division multiplexing (OFDM) has become a cornerstone modulation technique in modern communication systems due to its ability to transform a frequency-selective channel into a set of parallel flat-fading subchannels, which greatly simplifies equalization. As a result, OFDM has been widely adopted in standards such as Digital Audio Broadcasting (DAB) and wireless LANs. 

In OFDM systems, reliable channel estimation is crucial. For systems that employ pilots, channel estimation can be performed using classical least-squares (LS) and minimum mean-square error (MMSE) estimators. (e.g., \cite{Coleri2002, Proakis95}). Recently, advanced approaches that utilize deep neural networks \cite{Ye18, Ju24} have also been demonstrated to achieve excellent performance with data-driven methods. On the other hand, DAB-like systems inherently lack pilot symbols. Such systems use a differential encoder and a two-symbol differential detection (2SDD), effectively avoiding the need for phase estimation. While simple, 2SDD suffers from significant noise enhancement. Multi-symbol differential detection (MSDD) improves performance, and for many observations, its performance approaches that of ideal coherent detection \cite{Divsalar90, Colavolpe04}.

Although differential encoding was originally introduced for noncoherent detection, later research showed that its recursive structure can be exploited in a turbo construction, which achieves substantial gains through iterative processing \cite{Narayanan99, Hoeher99, CHC24_3, Brink00, WvH10}. While this turbo-differentially encoded (DE)-PSK construction offers significant performance gains by making the DE-PSK an inner code, it relies on accurate blind channel estimation, which is a key challenge in DAB-like scenarios.

Blind receivers with the EM algorithm can achieve near-optimal estimation in OFDM systems (e.g., \cite{Ma04, Liu14}). However, their performance strongly depends on the quality of the initial conditions \cite{CHC25}, and the EM iteration significantly increases system complexity. An alternative blind approach was introduced in \cite{Peleg00}, which proposed a trellis-based decomposition for joint channel phase and PSK symbol detection. Unlike MMSE or EM methods, this approach achieves maximum a-posteriori (MAP) optimality under fine phase quantization. The principle was later extended to block-fading channels with channel-gain estimation \cite{RRC03}, and further adapted to two-dimensional (2D) OFDM for DAB receivers \cite{WvH12}.

The main contribution of this paper is the generalization of the system in \cite{WvH12} into a fully blind turbo-DE-PSK receiver, and its extensive evaluation under realistic DAB-like system settings. We reformulate the joint phase estimation and detection problem as a unified hidden Markov model (HMM) and augment it with a noise variance estimator that exploits the null tones of the signal. Robustness is assessed through comprehensive evaluations under practical time-varying channels and standard transmission formats. In particular, we examine the effects of inner code length, phase quantization resolution, and 2D block size, providing insights into the fundamental trade-offs among receiver performance, channel coherence time, bandwidth, and complexity.


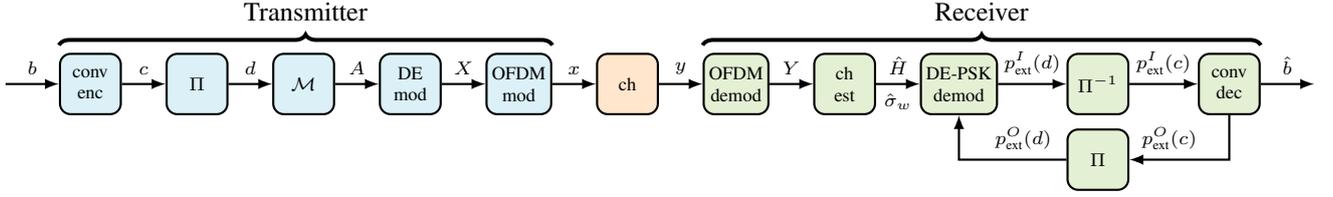
\begin{figure*}[t]
    \centering
    {\begin{tikzpicture}[line width=1.5pt, 
block1/.style={rectangle, fill=SkyBlue!20, text opacity=1, rounded corners, draw, inner sep=2pt, minimum width=8mm, minimum height=8mm, align=center, font=\scriptsize, thick},
linestyle/.style = {-latex,>={Latex[length=1mm,width=1mm]},thick},]

\tikzstyle{block2} = [block1, fill=orange!20]
\tikzstyle{block3} = [block1, fill=LimeGreen!20]

\def\dist{2.5}
\def\ttsize{\scriptsize}

\node[] at (0,\dist) (b) {};
\node[block1, right=0.23*\dist of b, align=center] (Inner) {conv\\enc};
\draw[linestyle] ([xshift=-20pt]Inner.west) -- (Inner.west) node[midway,above] {\ttsize{${b}$}};

\node[block1, right=0.23*\dist of Inner.east, align=center] (PI) {$\Pi$};
\draw[linestyle] (Inner.east) -- (PI.west) node[midway,above] {\ttsize{${c}$}};

\node[block1, right=0.23*\dist of PI.east, align=center] (Map) {$\mathcal{M}$};
\draw[linestyle] (PI.east) -- (Map.west) node[midway,above] {\ttsize{${d}$}};

\node[block1, right=0.23*\dist of Map.east, align=center] (DE) {DE\\mod};
\draw[linestyle] (Map.east) -- (DE.west) node[midway,above] {\ttsize{${A}$}};

\node[block1, right=0.23*\dist of DE.east, align=center] (ofdm) {OFDM\\mod};
\draw[linestyle] (DE.east) -- (ofdm.west) node[midway,above] {\ttsize{${X}$}};

\node[block2, right=0.23*\dist of ofdm.east, align=center] (CH) {ch};
\draw[linestyle] (ofdm.east) -- (CH.west) node[midway,above] {\ttsize{${x}$}};

\node[block3, right=0.23*\dist of CH.east, align=center] (dofdm) {OFDM\\demod};
\draw[linestyle] (CH.east) -- (dofdm.west) node[midway,above] {\ttsize{${y}$}};

\node[block3, right=0.23*\dist of dofdm.east, align=center] (est) {ch\\est};
\draw[linestyle] (dofdm.east) -- (est.west) node[midway,above] {\ttsize{$Y$}};

\node[block3, right=0.23*\dist of est.east, align=center] (dem) {DE-PSK\\demod};
\draw[linestyle] (est.east) -- (dem.west) node[midway,above] {\ttsize{$\hat{H}$}} node[midway,below] {\ttsize{$\hat{\sigma}_w$}};

\node[block3, right=0.36*\dist of dem.east, align=center] (invPI) {$\Pi^{-1}$};
\draw[linestyle] (dem.east) -- (invPI.west) node[midway,above] {\ttsize{$p^I_{\text{ext}}(d)$}};

\node[block3, right=0.36*\dist of invPI.east, align=center] (dec) {conv \\ dec};
\draw[linestyle] (invPI.east) -- (dec.west) node[midway,above] {\ttsize{$p^I_{\text{ext}}(c)$}};
\draw[linestyle] (dec.east) -- ([xshift=20pt]dec.east) node[midway,above] {\ttsize{${\hat{b}}$}};

\node[block3, below=0.07*\dist of invPI.south, align=center] (PI2) {$\Pi$};
\draw[linestyle] (dec.south) |- (PI2.east) node[pos=0.8,above] {\ttsize{$p^O_{\text{ext}}(c)$}};
\draw[linestyle] (PI2.west) -| (dem.south) node[pos=0.2,above] {\ttsize{$p^O_{\text{ext}}(d)$}};

\draw[decorate,decoration={brace,amplitude=4pt}]
  ($(Inner.north west)+(0,1mm)$) -- ($(ofdm.north east)+(0,1mm)$)
  node[midway,above=5pt] {\normalsize Transmitter};
\draw[decorate,decoration={brace,amplitude=4pt}] 
  ($(dofdm.north west)+(0,1mm)$) -- ($(dec.north east)+(0,1mm)$)
  node[midway,above=5pt] {\normalsize Receiver};


\end{tikzpicture}} 
    \caption{System block diagram of a convolutionally encoded, bit-interleaved PSK mapper with differential encoding and OFDM modulation at the transmitter side. The receiver comprises an OFDM demodulator, channel estimator, and an iterative demodulation/decoding module.}   
    \label{fig:sysblock}
\end{figure*}

\section{Differentially Encoded OFDM Transmission} \label{sec:sys_model}

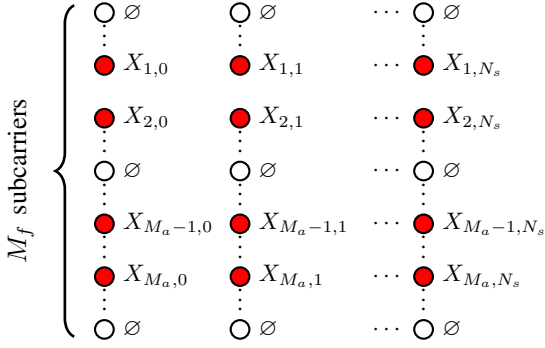
\begin{figure}[t]
    \centering
    {\begin{tikzpicture}[>=stealth]
\def\dx{2.1}   
\def\dy{0.7}   
\tikzset{
  nodebase/.style={circle,draw,thick,inner sep=0pt,minimum size=7pt},
  zero/.style={nodebase,fill=white},
  xnode/.style={nodebase,fill=red,inner sep=1pt},
  xlab/.style={anchor=west,scale=0.9}
}

\node[zero] (n-1-1) at (0*\dx,0*\dy) {};
\node[zero] (n-1-2) at (0.85*\dx,0*\dy) {};
\node[zero] (n-1-3) at (2*\dx,0*\dy) {};
\node[xlab, anchor = east] at (n-1-3.west) {$\cdots$};
\node[xlab] at (n-1-1.east) {$\varnothing$};
\node[xlab] at (n-1-2.east) {$\varnothing$};
\node[xlab] at (n-1-3.east) {$\varnothing$};

\node[xnode] (n-2-1) at (0*\dx,-1*\dy) {};
\node[xnode] (n-2-2) at (.85*\dx,-1*\dy) {};
\node[xnode] (n-2-3) at (2*\dx,-1*\dy) {};
\node[xlab,anchor=south] at (n-2-1.north) {$\vdots$};
\node[xlab,anchor=south] at (n-2-2.north) {$\vdots$};
\node[xlab,anchor=south] at (n-2-3.north) {$\vdots$};
\node[xlab] at (n-2-1.east) {$X_{1,0}$};
\node[xlab] at (n-2-2.east) {$X_{1,1}$};
\node[xlab] at (n-2-3.east) {$X_{1,N_s}$};
\node[xlab, anchor = east] at (n-2-3.west) {$\cdots$};

\node[xnode] (n-3-1) at (0*\dx,-2*\dy) {};
\node[xnode] (n-3-2) at (.85*\dx,-2*\dy) {};
\node[xnode] (n-3-3) at (2*\dx,-2*\dy) {};
\node[xlab] at (n-3-1.east) {$X_{2,0}$};
\node[xlab] at (n-3-2.east) {$X_{2,1}$};
\node[xlab, anchor = east] at (n-3-3.west) {$\cdots$};
\node[xlab] at (n-3-3.east) {$X_{2,N_s}$};

\node[zero] (n-4-1) at (0*\dx,-3*\dy) {};
\node[zero] (n-4-2) at (.85*\dx,-3*\dy) {};
\node[zero] (n-4-3) at (2*\dx,-3*\dy) {};
\node[xlab,anchor=south] at (n-4-1.north) {$\vdots$};
\node[xlab,anchor=south] at (n-4-2.north) {$\vdots$};
\node[xlab,anchor=south] at (n-4-3.north) {$\vdots$};
\node[xlab, anchor = east] at (n-4-3.west) {$\cdots$};
\node[xlab] at (n-4-1.east) {$\varnothing$};
\node[xlab] at (n-4-2.east) {$\varnothing$};
\node[xlab] at (n-4-3.east) {$\varnothing$};

\node[xnode] (n-5-1) at (0*\dx,-4*\dy) {};
\node[xnode] (n-5-2) at (.85*\dx,-4*\dy) {};
\node[xnode] (n-5-3) at (2*\dx,-4*\dy) {};
\node[xlab, anchor = east] at (n-5-3.west) {$\cdots$};
\node[xlab,anchor=south] at (n-5-1.north) {$\vdots$};
\node[xlab,anchor=south] at (n-5-2.north) {$\vdots$};
\node[xlab,anchor=south] at (n-5-3.north) {$\vdots$};
\node[xlab] at (n-5-1.east) {$X_{M_a-1,0}$};
\node[xlab] at (n-5-2.east) {$X_{M_a-1,1}$};
\node[xlab] at (n-5-3.east) {$X_{M_a-1,N_s}$};

\node[xnode] (n-6-1) at (0*\dx,-5*\dy) {};
\node[xnode] (n-6-2) at (.85*\dx,-5*\dy) {};
\node[xnode] (n-6-3) at (2*\dx,-5*\dy) {};
\node[xlab,anchor=south] at (n-6-1.north) {$\vdots$};
\node[xlab] at (n-6-1.east) {$X_{M_a,0}$};
\node[xlab,anchor=south] at (n-6-2.north) {$\vdots$};
\node[xlab,anchor=south] at (n-6-3.north) {$\vdots$};
\node[xlab, anchor = east] at (n-6-3.west) {$\cdots$};
\node[xlab] at (n-6-2.east) {$X_{M_a,1}$};
\node[xlab] at (n-6-3.east) {$X_{M_a,N_s}$};
\node[zero] (n-7-1) at (0*\dx,-6*\dy) {};
\node[zero] (n-7-2) at (.85*\dx,-6*\dy) {};
\node[zero] (n-7-3) at (2*\dx,-6*\dy) {};
\node[xlab,anchor=south] at (n-7-1.north) {$\vdots$};
\node[xlab,anchor=south] at (n-7-2.north) {$\vdots$};
\node[xlab,anchor=south] at (n-7-3.north) {$\vdots$};
\node[xlab, anchor = east] at (n-7-3.west) {$\cdots$};
\node[xlab] at (n-7-1.east) {$\varnothing$};
\node[xlab] at (n-7-2.east) {$\varnothing$};
\node[xlab] at (n-7-3.east) {$\varnothing$};
\draw[line width=1pt, decorate,decoration={brace,amplitude=7pt, mirror}]
  ($(n-1-1.north west)+(-3mm,0)$) -- ($(n-7-1.south west)+(-3mm,0)$)
  node[pos=0.25,left=20pt, rotate = 90] {\normalsize $M_f$ subcarriers};

\end{tikzpicture}} 
    \caption{DE-PSK symbol organization before IDFT. $\varnothing$ denotes the null tones used to avoid interference.}   
    \label{fig:2DX}
\end{figure}

The notation convention used in this paper is defined as follows. Time-domain quantities are written in lowercase, while frequency-domain quantities are written in uppercase. Scalars denote individual entries (e.g., $x_k$ and $X_m$). Vectors are denoted by bold italic letters. For example, a time-domain vector from index $1$ to $K$ is written as $\boldsymbol{x}_1^K=[x_1, x_2, \ldots, x_K]^\mathsf{T}$, and a vector extracted from the $m$-th row and $n$-th column of a matrix is denoted by $\boldsymbol{X}_m$ and $\boldsymbol{X}^{(n)}$. Matrices are denoted by bold uppercase letters (e.g., $\mathbf{X}$), with $X_{m,n}$ representing the $(m,n)$-th entry. Calligraphic letters denote sets, and $|\cdot|$ indicates their cardinality (e.g., $|\mathcal{X}|$). The transpose, Hermitian transpose, and complex conjugate are denoted by $(\cdot)^\mathsf{T}$, $(\cdot)^\mathsf{H}$, and $(\cdot)^\ast$, respectively. 

The transmitter is shown with light blue blocks in Fig.~\ref{fig:sysblock}. The information bit sequence $\boldsymbol{b}_1^{KR}$ is first processed by a convolutional encoder that produces the coded bit sequence $\boldsymbol{c}_1^{K}$. Here, $R$ represents the coding rate and $K$ is the total codeword length per transmission frame. A bit-level interleaver ($\Pi$) is used to permute the convolutional encoder output, where the interleaved coded bit sequence is represented as $\boldsymbol{d}_1^{K}$. Then, an $M$-PSK symbol mapper $\mathcal{M}$ maps the binary coded bits to symbol sequence $\boldsymbol{A}_1^T$ where $A_t \in \mathcal{X}=\{e^{j2\pi i /Q} \mid i=0,1,\dots,Q-1 \}$ with $Q=|\mathcal{X}|$ represents the cardinality of the PSK symbol space.

In an OFDM system, the information $\boldsymbol{A}_1^T$ is encoded in the active subcarriers. Let $M_a$ be the number of active subcarriers and $N_s$ the number of OFDM symbols. The 1D PSK symbol vector $\boldsymbol{A}_1^T$ is reshaped to a 2D matrix $\mathbf{A}\in\mathbb{C}^{M_a\times N_s}$, where row $m$ represents the $m$-th subcarrier and column $n$ denotes the $n$-th time-domain OFDM symbol. In our system design, a differential symbol encoder modulates the incoming PSK symbols and generates the DE-PSK symbol sequence per subcarrier $\boldsymbol{X}_m=[X_{m,0},X_{m,1},\cdots,X_{m,N_s}]$ via 
\begin{align} \label{eq:diff_mod}
    X_{m,n} = A_{m,n} \cdot X_{m,n-1},    
\end{align}
where we define $X_{m,0} = 1$ and $X_{m,n} \in \mathcal{X}$. 

After mapping the DE-PSK symbols to the active subcarriers, $M_n$ null tones (such as guard bands and DC subcarriers) are added to prevent interference, resulting in a total subcarrier size of $M_f$. A schematic for the 2D DE-PSK symbol construction for OFDM processing is shown in Fig.~\ref{fig:2DX}. The OFDM modulator then performs the inverse discrete Fourier transform (IDFT) to generate the time-domain OFDM symbols via $\mathbf{x} = \mathbf{F}^H \mathbf{X}$, where $\mathbf{F}^H$ is the $M_f$-point IDFT matrix. We then append $N_{cp}$ cyclic prefix symbols in front of each OFDM symbol to prevent inter-symbol interference. Finally, the appended 2D time-domain matrix is converted back to a 1D vector $\boldsymbol{x}_1^{M_f(N_{cp}+N_s+1)}$ for transmission.

We consider a frequency-selective channel where the input-output relationship can be expressed as $y_t = \sum_{i=0}^{I-1} h_{i,t}\, x_{t-i} + w_t$ where $h_{i,t}$ denotes the time-varying channel coefficient of the $i$-th tap at time $t$. 
The noise realizations $w_t \sim \mathcal{CN}(0,\sigma_w^2)$ are modeled as circularly symmetric complex Gaussian random variables with zero mean and variance $\sigma_w^2$. 

After cyclic prefix removal, each OFDM symbol of length $M_f$ is reshaped into a vector $\boldsymbol{y}^{(n)}=[y_{1,n},y_{2,n},\cdots,y_{M_f,n}]^\mathsf{T}$, where $n=0,1,\ldots,N_s$ denotes the block index. 
Applying DFT, represented abstractly by the unitary matrix, yields
\begin{align} \label{eq:ch_freq}
    \boldsymbol{Y}^{(n)} = \mathbf{F} \boldsymbol{y}^{(n)} 
    = \boldsymbol{H}^{(n)} \odot \boldsymbol{X}^{(n)} + \boldsymbol{W}^{(n)},
\end{align}
where $\odot$ denotes element-wise multiplication, $\boldsymbol{W}^{(n)}$ is the transformed noise vector with entries $W_{m,n}\sim\mathcal{CN}(0,\sigma_w^2)$, and $\boldsymbol{H}^{(n)}$ denote the channel frequency response vector. From \eqref{eq:ch_freq}, the time-domain channel convolution simplifies to element-wise multiplication. The channel frequency response can be further expressed as
\begin{align}\label{eq:ch}
    H_{m,n} =G_{m,n} \, e^{j\phi_{m,n}},
\end{align}
where $G_{m,n}= | H_{m,n} |$ and $\phi_{m,n}$ denote the element-wise channel gain and phase, respectively.

\section{Optimal DE-PSK Demodulation and Decoding}\label{sec:opt_rec}
The turbo-DE-PSK receiver design is shown with light green blocks in Fig.~\ref{fig:sysblock}. This section first introduces the optimal demodulation scheme under the assumption of perfect channel knowledge \eqref{eq:ch}. We will outline the BCJR algorithm for MAP demodulation and its extension to an iterative inference system for coded DE-PSK modulation. In the following derivation, the mathematical expressions are identical across all subcarriers $m$ and the operations are performed along the OFDM symbol index $n$. Therefore, we omit $m$ for notational simplicity (e.g., we write $Y_n$ in place of $Y_{m,n}$).

\subsection{HMM and the BCJR algorithm in OFDM}
An HMM is a statistical model that captures the temporal behavior of the observable process $\boldsymbol{Y}$ whose outcomes depend on the corresponding latent variables $\boldsymbol{S}$. At any specific time, the latent variable $S_{n}$ is drawn from a time-invariant finite-state set $\mathcal{S} \triangleq \{0,1,\cdots,|\mathcal{S}|-1\}$, where $|\mathcal{S}|$ is the size of the state space. The evolution of the states follows a Markov process with transition probabilities $p(S_{n} = j \mid S_{n-1}=i)$, which describes the probability of moving from state $i$ at OFDM time index $n-1$ to state $j$ at time instants $n$. The relationship between the latent state and the observation is described by the observation likelihood \( p(Y_{n} \mid S_{n}=j) \), which indicates the probability of an observation being generated from state $j$ at OFDM time index $n$.

The Markovian structure of an HMM enables the efficient detection of hidden states through the well-known BCJR algorithm \cite{bcjr74}. Specifically, the inference problem aims at calculating the posterior belief of a latent state $S_n$ at a specific instance given the sequence of observations on a specific subcarrier $\boldsymbol{Y}_0^{N_s}$.
\begin{align} 
    p(S_{n} \mid \boldsymbol{Y}_0^{N_s}) &\propto p(S_{n}, \boldsymbol{Y}_0^{N_s}) = \sum_{S_{n-1}\in\mathcal{S}}{p(S_{n}, S_{n-1}, \boldsymbol{Y}_0^{N_s})}, \nonumber
\end{align}
where the joint probability on the right-hand side can be further factorized as
\begin{align} 
    p(S_{n}, &S_{n-1},\boldsymbol{Y}_0^{N_s}) = {\alpha(S_{n-1})} \cdot {\gamma(S_{n-1},S_{n})}  \cdot {\beta(S_{n})}. \label{eq:bcjr_joi} 
\end{align}

Here, $\alpha$, $\beta$, and $\gamma$ denote the forward recursion, backward recursion, and the branch metric, respectively, which can be further derived as
\begin{align}
    &\alpha(S_{n}) = \textstyle\sum_{S_{n-1}\in\mathcal{S}} \gamma(S_{n}, S_{n-1}) \cdot \alpha(S_{n-1}), \label{eq:siso_alpha}  \\ 
    &\beta(S_{n-1}) = \textstyle\sum_{S_{n}\in\mathcal{S}} \beta(S_n) \cdot \gamma(S_{n}, S_{n-1}),  \label{eq:siso_beta} \\
    &\gamma(S_{n-1}, S_{n}) = p(Y_n| S_n) \cdot p(S_n|S_{n-1}). \label{eq:gamma}
\end{align}
The full derivation of the BCJR algorithm can be found in \cite{bcjr74}, and its variant for MAP-based turbo-DE-PSK demodulation and decoding is detailed in \cite{CHC24_3}.

\begin{figure*}[t]
  \centering
  \begin{subfigure}[t]{0.3\textwidth}
    \centering
    \usetikzlibrary{positioning}

\def\dx{0.8}   
\def\dy{1.7}   
\def\x0{0}     
\def\y0{0}     
\def\lblsep{1mm} 

\tikzstyle{state}=[shape=circle,draw=black!50,fill=gray!50,font=\scriptsize, minimum size = 18pt]
\tikzstyle{mainstate}=[state,very thick]
\tikzstyle{edge}=[-stealth, semithick]  
\tikzstyle{edge1}=[edge, Black]
\tikzstyle{edge2}=[edge, Cyan]
\tikzstyle{edge3}=[edge, Orange]
\tikzstyle{edge4}=[edge, Green]

\begin{tikzpicture}[x=1cm,y=1cm]

\node[mainstate] (s1_top) at ({\x0+(-1.5)*\dx}, {\y0+0}) {$0$};
\node[mainstate] (s2_top) at ({\x0+(-0.5)*\dx}, {\y0+0}) {$1$};
\node[mainstate] (s3_top) at ({\x0+( 0.5)*\dx}, {\y0+0}) {$2$};
\node[mainstate] (s4_top) at ({\x0+( 1.5)*\dx}, {\y0+0}) {$3$};

\node[mainstate] (s1_bot) at ({\x0+(-1.5)*\dx}, {\y0-\dy}) {$0$};
\node[mainstate] (s2_bot) at ({\x0+(-0.5)*\dx}, {\y0-\dy}) {$1$};
\node[mainstate] (s3_bot) at ({\x0+( 0.5)*\dx}, {\y0-\dy}) {$2$};
\node[mainstate] (s4_bot) at ({\x0+( 1.5)*\dx}, {\y0-\dy}) {$3$};

\node[left=2mm of s1_top] {\small $S_{n-1}$};
\node[left=2mm of s1_bot] {\small $S_n$};

\node[anchor=south] at ([yshift=\lblsep]s1_top.north) {\scriptsize $e^{j0}$};
\node[anchor=south] at ([yshift=\lblsep]s2_top.north) {\scriptsize $e^{j\frac{\pi}{2}}$};
\node[anchor=south] at ([yshift=\lblsep]s3_top.north) {\scriptsize $e^{j\pi}$};
\node[anchor=south] at ([yshift=\lblsep]s4_top.north) {\scriptsize $e^{j\frac{3\pi}{2}}$};

\path
  (s1_top.south) edge[edge1] (s1_bot.north)
           edge[edge2] (s2_bot.north)
           edge[edge3] (s3_bot.north)
           edge[edge4] (s4_bot.north)
  (s2_top.south) edge[edge4] (s1_bot.north)
           edge[edge1] (s2_bot.north)
           edge[edge2] (s3_bot.north)
           edge[edge3] (s4_bot.north)
  (s3_top.south) edge[edge3] (s1_bot.north)
           edge[edge4] (s2_bot.north)
           edge[edge1] (s3_bot.north)
           edge[edge2] (s4_bot.north)
  (s4_top.south) edge[edge2] (s1_bot.north)
           edge[edge3] (s2_bot.north)
           edge[edge4] (s3_bot.north)
           edge[edge1] (s4_bot.north);


\end{tikzpicture}
    \caption{Optimal DE-QPSK demodulator}
    \label{fig:trellis_dqpsk}
  \end{subfigure}
  \begin{subfigure}[t]{0.65\textwidth}
    \centering
    \usetikzlibrary{positioning}

\def\dx{0.7}   
\def\dy{1.7}   
\def\x0{0}     
\def\y0{0}     
\def\lblsep{1mm} 

\tikzstyle{state}=[shape=circle,draw=black!50,fill=gray!50,font=\scriptsize, minimum size = 18pt ]
\tikzstyle{mainstate}=[state,very thick]
\tikzstyle{edge}=[-stealth, semithick]  
\tikzstyle{edge1}=[edge, Black]
\tikzstyle{edge2}=[edge, Cyan]
\tikzstyle{edge3}=[edge, Orange]
\tikzstyle{edge4}=[edge, Green]
\tikzstyle{edge12}=[edge,thick,  Black, dotted]
\tikzstyle{edge22}=[edge,thick,  Cyan, dotted]
\tikzstyle{edge32}=[edge,thick,  Orange, dotted]
\tikzstyle{edge42}=[edge,thick,  Green, dotted]

\begin{tikzpicture}[x=1cm,y=1cm]

\node[mainstate] (s1_top) at ({\x0+(-1.5)*\dx}, {\y0+0}) {$0$};
\node[mainstate] (s2_top) at ({\x0+(-0.5)*\dx}, {\y0+0}) {$1$};
\node[] (s3_top) at ({\x0+(0.5)*\dx}, {\y0+0}) {$\cdots$};
\node[mainstate] (s4_top) at ({\x0+(1.5)*\dx}, {\y0+0}) {$8$};
\node[mainstate] (s5_top) at ({\x0+(2.5)*\dx}, {\y0+0}) {$9$};
\node[] (s6_top) at ({\x0+(3.5)*\dx}, {\y0+0}) {$\cdots$};
\node[mainstate] (s7_top) at ({\x0+( 4.5)*\dx}, {\y0+0}) {$16$};
\node[mainstate] (s8_top) at ({\x0+( 5.5)*\dx}, {\y0+0}) {$17$};
\node[] (s9_top) at ({\x0+(6.5)*\dx}, {\y0+0}) {$\cdots$};
\node[mainstate] (s10_top) at ({\x0+( 7.5)*\dx}, {\y0+0}) {$24$};
\node[mainstate] (s11_top) at ({\x0+( 8.5)*\dx}, {\y0+0}) {$25$};
\node[] (s12_top) at ({\x0+(9.5)*\dx}, {\y0+0}) {$\cdots$};

\node[mainstate] (s1_bot) at ({\x0+(-1.5)*\dx}, {\y0-\dy}) {$0$};
\node[mainstate] (s2_bot) at ({\x0+(-0.5)*\dx}, {\y0-\dy}) {$1$};
\node[] (s3_bot) at ({\x0+(0.5)*\dx}, {\y0-\dy}) {$\cdots$};
\node[mainstate] (s4_bot) at ({\x0+(1.5)*\dx}, {\y0-\dy}) {$8$};
\node[mainstate] (s5_bot) at ({\x0+(2.5)*\dx}, {\y0-\dy}) {$9$};
\node[] (s6_bot) at ({\x0+(3.5)*\dx}, {\y0-\dy}) {$\cdots$};
\node[mainstate] (s7_bot) at ({\x0+( 4.5)*\dx}, {\y0-\dy}) {$16$};
\node[mainstate] (s8_bot) at ({\x0+( 5.5)*\dx}, {\y0-\dy}) {$17$};
\node[] (s9_bot) at ({\x0+(6.5)*\dx}, {\y0-\dy}) {$\cdots$};
\node[mainstate] (s10_bot) at ({\x0+( 7.5)*\dx}, {\y0-\dy}) {$24$};
\node[mainstate] (s11_bot) at ({\x0+( 8.5)*\dx}, {\y0-\dy}) {$25$};
\node[] (s12_bot) at ({\x0+(9.5)*\dx}, {\y0-\dy}) {$\cdots$};

\node[left=2mm of s1_top] {\small $\tilde{S}_{n-1}$};
\node[left=2mm of s1_bot] {\small $\tilde{S}_n$};

\node[anchor=south] at ([yshift=\lblsep]s1_top.north) {\scriptsize $e^{j0}$};
\node[anchor=south] at ([yshift=\lblsep]s2_top.north) {\scriptsize $e^{j\frac{\pi}{16}}$};
\node[anchor=south] at ([yshift=\lblsep]s4_top.north) {\scriptsize $e^{j\frac{\pi}{2}}$};
\node[anchor=south] at ([yshift=\lblsep]s5_top.north) {\scriptsize $e^{j\frac{9\pi}{16}}$};
\node[anchor=south] at ([yshift=\lblsep]s7_top.north) {\scriptsize $j\pi$};
\node[anchor=south] at ([yshift=\lblsep]s8_top.north) {\scriptsize $e^{j\frac{17\pi}{16}}$};
\node[anchor=south] at ([yshift=\lblsep]s10_top.north) {\scriptsize $e^{j\frac{3\pi}{2}}$};
\node[anchor=south] at ([yshift=\lblsep]s11_top.north) {\scriptsize $e^{j\frac{25\pi}{16}}$};
\path
  (s1_top.south) edge[edge1] (s1_bot.north)
           edge[edge2] (s4_bot.north)
           edge[edge3] (s7_bot.north)
           edge[edge4] (s10_bot.north)
  (s2_top.south) edge[edge12] (s2_bot.north)
           edge[edge22] (s5_bot.north)
           edge[edge32] (s8_bot.north)
           edge[edge42] (s11_bot.north)
  (s4_top.south) edge[edge4] (s1_bot.north)
           edge[edge1] (s4_bot.north)
           edge[edge2] (s7_bot.north)
           edge[edge3] (s10_bot.north)
  (s5_top.south) edge[edge42] (s2_bot.north)
           edge[edge12] (s5_bot.north)
           edge[edge22] (s8_bot.north)
           edge[edge32] (s11_bot.north)
  (s7_top.south) edge[edge3] (s1_bot.north)
           edge[edge4] (s4_bot.north)
           edge[edge1] (s7_bot.north)
           edge[edge2] (s10_bot.north)
  (s8_top.south) edge[edge32] (s2_bot.north)
           edge[edge42] (s5_bot.north)
           edge[edge12] (s8_bot.north)
           edge[edge22] (s11_bot.north)
  (s10_top.south) edge[edge2] (s1_bot.north)
           edge[edge3] (s4_bot.north)
           edge[edge4] (s7_bot.north)
           edge[edge1] (s10_bot.north)
  (s11_top.south) edge[edge22] (s2_bot.north)
           edge[edge32] (s5_bot.north)
           edge[edge42] (s8_bot.north)
           edge[edge12] (s11_bot.north);

\node[anchor=west, fill=white, inner sep=0.1pt, font=\footnotesize, yshift=-7mm, xshift=0mm]
  at (s12_top.east) {
  \begin{tabular}{@{}c@{}c@{}}
    \multicolumn{2}{c}{Trellis Input} \\
    \begin{tikzpicture}\draw[edge1,-stealth] (0.2,0) -- (.6,0);\end{tikzpicture} & $e^{j0}$ \\
    \begin{tikzpicture}\draw[edge2,-stealth] (0.2,0) -- (.6,0);\end{tikzpicture} & $e^{j\frac{\pi}{2}}$ \\
    \begin{tikzpicture}\draw[edge3,-stealth] (0.2,0) -- (.6,0);\end{tikzpicture} & $e^{j\pi}$ \\
    \begin{tikzpicture}\draw[edge4,-stealth] (0.2,0) -- (.6,0);\end{tikzpicture} & $e^{j\frac{3\pi}{2}}$ \\
  \end{tabular}
};

\end{tikzpicture}
    \caption{Joint phase and DE-QPSK demodulator}
    \label{fig:trellis_phs}
  \end{subfigure}
  \caption{Trellis constructions for the (a) optimal DQPSK symbol demodulator and the (b) joint channel phase and DQPSK symbol demodulator with $L=32$.}
  \label{fig:trellis_overall}
\end{figure*}
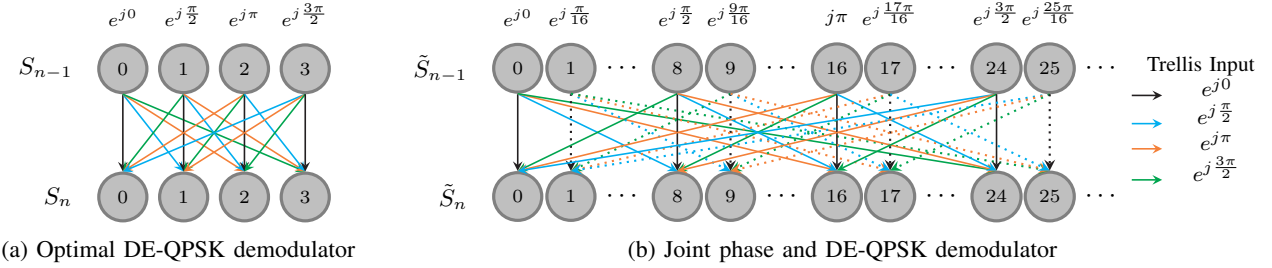


\subsection{MAP DE-PSK Demodulator}
To design the MAP demodulator for DE-PSK symbols, we first define the latent state as
\begin{align} 
    S_n \triangleq f(X_n \in\mathcal{X}) \in \mathcal{S} \triangleq \{0,1,\ldots,Q-1\}, \nonumber
\end{align}
where $f(\cdot)$ is a mapping function that maps each input symbol to a unique discrete latent state. As the state realization $S_n$ is uniquely determined by the DE-PSK symbol $X_n$, the state transition is driven by the PSK-modulated symbol $A_n$, which captures the structure imposed by the differential encoder in \eqref{eq:diff_mod}. This relation is expressed through the state-transition probability
\begin{align} \label{eq:P_tran1}
    p(S_n \mid S_{n-1}) = p(A_n), \quad A_n: S_{n-1} \rightarrow S_n,
\end{align}
where $ A_n: S_{n-1} \rightarrow S_n$ indicates that the transition from $S_{n-1}$ to $S_n$ is driven by the PSK symbol $A_n$. In Fig.~\ref{fig:trellis_dqpsk}, we show the trellis construction for an ideal DE quadrature PSK (QPSK) demodulator with the colored connections indicating the state transitions driven by specific trellis input $A_n$. The observation likelihood given the knowledge of the channel response can be written as
\begin{align}\label{eq:lik_ideal}
     p(Y_n\mid S_n, H_n) = \frac{1}{{\pi \sigma_w^2}}\exp{\left(-\frac{|Y_n-H_nX_n|^2}{\sigma^2_w}\right)}.
\end{align}

We can now rewrite \eqref{eq:gamma} and express the trellis branch metric for the DE-PSK symbol demodulator based on \eqref{eq:P_tran1} and \eqref{eq:lik_ideal} as 
\begin{align} \label{eq:turbo_gamma1}
    \gamma(S_n, S_{n-1}) \triangleq p(Y_n\mid S_n,H_n) \cdot p(A_n) , \quad A_n:S_{n-1} \rightarrow S_n.
\end{align}
The joint probability of the transmitted PSK symbol is then obtained by 
\begin{align}
    p(A_n, \boldsymbol{Y}_0^{N_s}) = \textstyle\sum_{A_n:S_{n-1}\rightarrow S_n}{ {p(S_n,S_{n-1}, \boldsymbol{Y}_0^{N_s})} }, \nonumber
\end{align}
where the joint probability $p(S_n, S_{n-1}, \boldsymbol{Y}_0^{N_s})$ is obtained via \eqref{eq:bcjr_joi}--\eqref{eq:siso_beta}.

Note that, under an ideal coherent demodulation scenario, the trellis is assumed to start in a known initial state $S_0=0$ and terminate uniformly across $Q$ possible states. We therefore set the initialization value as $\alpha(S_0)=\delta_{S_0,0}$ and $\beta(S_{N_s}) = 1/Q$ with $\delta_{i,j}$ denotes the Kronecker delta.

\subsection{Iterative Demodulation and Decoding}
As shown in Fig.~\ref{fig:sysblock}, after DE-PSK demodulation, the soft symbol information is first converted to bit information via
\begin{align} \label{eq:turbo_s2b}
    p(d_k, \boldsymbol{Y}_0^{N_s}) = \mathcal{M}^{-1}[p(A_n,\boldsymbol{Y}_0^{N_s})],
\end{align}
where $\mathcal{M}^{-1}$ denotes the inverse symbol mapping function that converts the PSK symbol probability back to the corresponding coded bit probability. Then, the extrinsic bit information from the demodulator can be computed via $p^I_{\text{ext}}(d_{k}) \triangleq {p(d_{k}, \boldsymbol{Y}_0^{N_s})} /{p^O_{\text{ext}}(d_{k})}$. Here, $p^I_{\text{ext}}(\cdot)$ and $p^O_\text{ext}(\cdot)$ denote the extrinsic information from the inner code (DE-PSK demodulation) and outer code (convolutional decoder), respectively.
After performing the bit-deinterleaving, we can derive the extrinsic likelihood information of the coded bits as 
\begin{align}\label{eq:ext_c}
     p^I_{\text{ext}}(c_{k}) = \Pi^{-1} \left[ p^I_{\text{ext}}(d_k) \right].
\end{align}

The convolutional decoder employs \eqref{eq:ext_c} and applies \eqref{eq:bcjr_joi}--\eqref{eq:gamma} to compute the joint probabilities of the information bits $p(b_k, \boldsymbol{Y}_0^{N_s}) $ and the coded bit $p(c_k, \boldsymbol{Y}_0^{N_s})$. As the decoder is not the main focus of this work, we will omit the detailed derivations here. Interested readers may refer to \cite{CHC24_3} for a comprehensive derivation. To compute the extrinsic prior information produced by the convolutional decoder, we divide the joint probability $p(c_k, \boldsymbol{Y}_0^{N_s})$ by the extrinsic information of the likelihood from the demodulator \( p^I_{\text{ext}}(c_k) \) as
$p^O_{\text{ext}}(c_k) \triangleq {p(c_k,\boldsymbol{Y}_0^{N_s})} / {p^I_{\text{ext}}(c_k)}$.
After the bit interleaver $p^O_{\text{ext}}(d_k) = \Pi \left[ p^O_{\text{ext}}(c_k) \right]$, we can derive the extrinsic prior information of the PSK symbol via bit-to-symbol mapping $p^O_{\text{ext}}(A_n) = \mathcal{M}[p^O_{\text{ext}}(d_{k})]$ which is then fed back as the prior information in \eqref{eq:turbo_gamma1} of the DE-PSK symbol demodulator for the next turbo iteration.

\section{Joint Blind Channel Estimation and Symbol Demodulation} \label{sec:rec_design}

In a blind OFDM system, the receiver lacks knowledge of the channel's phase and gain. Consequently, the optimal coherent demodulator that assumes a known phase (i.e., a MAP symbol demodulator on a trellis with $Q$ states) cannot be used directly. In this section, we reformulate the trellis decomposition method proposed in \cite{Peleg00, RRC03,WvH12} that jointly estimates the channel phase and detects the transmitted symbols. To estimate the channel gain, we utilize the power-averaged method proposed in \cite{RRC03}, and we extend this method to estimate noise variance as well by utilizing null tone symbols. This approach ultimately enables the development of a completely blind OFDM coherent receiver.

\subsection{Joint Phase Estimation and Symbol Demodulation}
To tackle the unknown channel phase, the unknown phase is discretized \cite{Peleg00} into $L$ equispaced hypotheses $\tilde{\phi} \in  \{2\pi l / L, l=0,1,\cdots,L-1\}$, where $L$ is an integer multiple of $Q$. These hypotheses are then embedded into the trellis, under the assumption that the channel phase is frame-wise invariant. The resulting MAP detection operates jointly on information symbols and phase states, effectively tracking the phase as part of the state evolution. We now define the latent state for our decomposed trellis as
\begin{align}
    \tilde{S}_n \triangleq f(X_n e^{j\phi}= e^{j\tilde{\phi}}) \in \tilde{\mathcal{S}} \triangleq \{0,1,\ldots,L -1\}, \nonumber
\end{align}
where the state transition probability
\begin{align} \label{eq:Ptran_est}
    p(\tilde{S}_n \mid \tilde{S}_{n-1}) = p(A_n), \quad A_n: \tilde{S}_{n-1} \rightarrow \tilde{S}_n,
\end{align}
and the channel likelihood 
\begin{align} \label{eq:lik_est}
     p(Y_n \mid \tilde{S}_n) = \frac{1}{{\pi \sigma_w^2}}\exp{\left(-\frac{|Y_n - G_n e^{j\tilde{\phi}}|^2}{\sigma^2_w}\right)}
\end{align}
are defined based on the state $\tilde{S}_n$ of the decomposed trellis.
Compared to \eqref{eq:lik_ideal}, we now incorporate both the discretized channel phase and the transmitted DE-PSK symbol into specific states. This extended trellis for joint phase and PSK symbol demodulation is shown in Fig.~\ref{fig:trellis_phs}. Given \eqref{eq:Ptran_est} and \eqref{eq:lik_est}, the branch metric of the decomposed trellis can be defined as
\begin{align}
    \gamma(\tilde{S}_n, \tilde{S}_{n-1}) \triangleq p(Y_n \mid \tilde{S}_n) \cdot p(A_n) , \quad A_n:\tilde{S}_{n-1} \rightarrow \tilde{S}_n. \nonumber
\end{align}
Following a process analogous to \eqref{eq:bcjr_joi}--\eqref{eq:siso_beta}, we can derive the joint state pair probability $p(\tilde{S}_n, \tilde{S}_{n-1}, \boldsymbol{Y}_0^{N_s})$ and obtain the joint symbol probability via marginalization  
\begin{align} \label{eq:map_a}
    p(A_n, \boldsymbol{Y}_0^{N_s}) = \textstyle\sum_{A_n:\tilde{S}_{n-1}\rightarrow \tilde{S}_n}{ {p(\tilde{S}_n,\tilde{S}_{n-1}, \boldsymbol{Y}_0^{N_s})} }. 
\end{align}
In the blind scenario, the first reference symbol is unknown. Therefore, the trellis is initialized with $\alpha(\tilde{S}_0)=p(Y_0|\tilde{S}_0)$ and $\beta(\tilde{S}_{N_s}) = 1/L$.

\begin{figure}[t]
    \centering
    {\begin{tikzpicture}[>=stealth]
\def\dx{2.1}   
\def\dy{0.7}   
\tikzset{
  nodebase/.style={circle,draw,thick,inner sep=0pt,minimum size=7pt},
  zero/.style={nodebase,fill=white},
  xnode/.style={nodebase,fill=red,inner sep=1pt},
  xlab/.style={anchor=west,scale=0.9}
}

\node[xnode] (n-2-1) at (0*\dx,-1*\dy) {};
\node[xnode] (n-2-2) at (.85*\dx,-1*\dy) {};
\node[xnode] (n-2-3) at (2*\dx,-1*\dy) {};
\node[xlab] at (n-2-1.east) {$Y_{1,0}$};
\node[xlab] at (n-2-2.east) {$Y_{1,1}$};
\node[xlab] at (n-2-3.east) {$Y_{1,N_s}$};
\node[xlab, anchor = east] at (n-2-3.west) {$\cdots$};

\node[xnode] (n-3-1) at (0*\dx,-2*\dy) {};
\node[xnode] (n-3-2) at (.85*\dx,-2*\dy) {};
\node[xnode] (n-3-3) at (2*\dx,-2*\dy) {};
\node[xlab] at (n-3-1.east) {$Y_{2,0}$};
\node[xlab] at (n-3-2.east) {$Y_{2,1}$};
\node[xlab, anchor = east] at (n-3-3.west) {$\cdots$};
\node[xlab] at (n-3-3.east) {$Y_{2,N_s}$};

\node[xnode] (n-5-1) at (0*\dx,-3.5*\dy) {};
\node[xnode] (n-5-2) at (.85*\dx,-3.5*\dy) {};
\node[xnode] (n-5-3) at (2*\dx,-3.5*\dy) {};
\node[xlab, anchor = east] at (n-5-3.west) {$\cdots$};
\node[xlab,anchor=south] at (n-5-1.north) {$\vdots$};
\node[xlab,anchor=south] at (n-5-2.north) {$\vdots$};
\node[xlab,anchor=south] at (n-5-3.north) {$\vdots$};
\node[xlab] at (n-5-1.east) {$Y_{M_a-1,0}$};
\node[xlab] at (n-5-2.east) {$Y_{M_a-1,1}$};
\node[xlab] at (n-5-3.east) {$Y_{M_a-1,N_s}$};

\node[xnode] (n-6-1) at (0*\dx,-4.5*\dy) {};
\node[xnode] (n-6-2) at (.85*\dx,-4.5*\dy) {};
\node[xnode] (n-6-3) at (2*\dx,-4.5*\dy) {};
\node[xlab,anchor=south] at (n-6-1.north) {$\vdots$};
\node[xlab] at (n-6-1.east) {$Y_{M_a,0}$};
\node[xlab,anchor=south] at (n-6-2.north) {$\vdots$};
\node[xlab,anchor=south] at (n-6-3.north) {$\vdots$};
\node[xlab, anchor = east] at (n-6-3.west) {$\cdots$};
\node[xlab] at (n-6-2.east) {$Y_{M_a,1}$};
\node[xlab] at (n-6-3.east) {$Y_{M_a,N_s}$};
\draw[line width=1pt, decorate,decoration={brace,amplitude=7pt, mirror}]
  ($(n-2-1.north west)+(-3mm,0)$) -- ($(n-6-1.south west)+(-3mm,0)$)
  node[pos=0.1,left=20pt, rotate = 90] {\normalsize $M_a$ subcarriers};

\node[] (x1) at (0*\dx,-1*\dy) {};
\node[] (x2) at (1.3*\dx,-2.55*\dy) {};
\node[draw, red, dashed, line width=1pt, rounded corners=2pt,
      fit=(x1)(x2),
     ] (Sel) {};

\node[red, above=0mm of Sel.north] {$N$};   
\node[red, right=0mm of Sel.east]  {$M$};   

\end{tikzpicture}} 
    \caption{Received signal after DFT, null tone removal, and cyclic prefix removal. The red rectangular box indicates the 2D blocks used for channel estimation and demodulation.}   
    \label{fig:2dsym}
\end{figure}
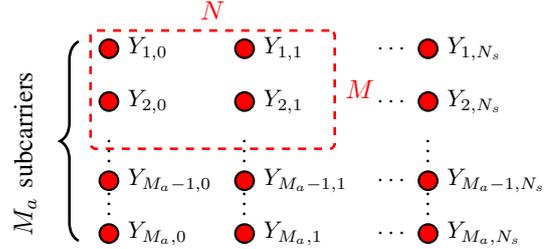

\subsection{2D Trellis Decomposition} \label{sec:sub-2d}
A closer look at the trellis in Fig.~\ref{fig:trellis_phs} shows that it decomposes into $L/Q$ mutually disconnected sections (sub-trellises) of size $Q$ (e.g., solid and dotted connecting lines in Fig.~\ref{fig:trellis_phs}) as identified in \cite{RRC03, WvH12}. These sub-trellises capture the channel-phase hypotheses and enable a natural factorization of the detection metric into (i) a \emph{phase-estimation} component that selects among the phase hypotheses, and (ii) a \emph{symbol-demodulation} component that performs MAP detection within the chosen phase sub-trellis.

To formalize this decomposition, \cite{WvH12} introduce a hidden sub-trellis index  $\tau \in \mathcal{T} \triangleq \{0,1,\ldots,L/Q-1\}$.
Each $\tau$ corresponds to a sub-trellis consisting of a subset of states from $\tilde{\mathcal{S}}$, defined as
\begin{align}\label{eq:subtrellis_def}
    \tilde{\mathcal{S}}_\tau 
    &\triangleq \Big\{\, l \;\big|\; l = \tau + p\frac{L}{Q}, \; p=0,1,\ldots,Q-1 \,\Big\} \subseteq \tilde{\mathcal{S}}.
\end{align}
The overall state space is covered by $\tilde{\mathcal{S}} = \bigcup_{\tau \in \mathcal{T}} \tilde{\mathcal{S}}_\tau$, with each $\tilde{\mathcal{S}}_\tau$ forming one independent sub-trellis. Using this partition, \eqref{eq:map_a} can be rewritten as \cite[Eq.~(37)]{WvH12}
\begin{align} \label{eq:map_sub}
    p(A_n, \boldsymbol{Y}_0^{N_s}) \propto \sum_{\tau \in \mathcal{T}} 
      p(\tau \mid \boldsymbol{Y}_0^{N_s}) \cdot p(A_n \mid \boldsymbol{Y}_0^{N_s}, \tau).
\end{align}
This expression highlights the factorization: phase estimation corresponds to selecting the correct sub-trellis (first term), while symbol demodulation is performed within the chosen $\tau$ (second term).

In DAB-like transmission, the length of the OFDM symbol $N_s$ is often limited, which negatively affects the performance of phase estimation. Furthermore, when the receiver operates under a rapidly changing channel, the coherence time restricts the number of available OFDM symbols for phase estimation. Therefore, to further improve phase recovery, we exploit the adjacent subcarriers with length $M$ and extend the trellis along the frequency dimension. We can now rewrite \eqref{eq:map_sub} with 2D trellis extension as 
\begin{align}
    p(&A_{m,n}, \boldsymbol{Y}_1,\boldsymbol{Y}_2,\cdots,\boldsymbol{Y}_M) \nonumber \\
    & \propto \sum_{\tau \in \mathcal{T}} p(\tau \mid \boldsymbol{Y}_1,\boldsymbol{Y}_2,\cdots,\boldsymbol{Y}_M) \cdot p(A_{m,n} \mid \boldsymbol{Y}_m, \tau), \label{eq:map_sub2D}
\end{align}
with $\boldsymbol{Y}_m=(Y_{m,0},Y_{m,1},\cdots,Y_{m,N})$ denoting the received sequence used per subcarrier.
In \eqref{eq:map_sub2D}, the first term on the right-hand side carries out phase estimation on the extended trellis, whereas the second term is identical to \eqref{eq:map_sub} because the PSK information is encoded independently per subcarrier. Moreover, the temporal span changes from $N_s$ in \eqref{eq:map_sub} to $N$ in \eqref{eq:map_sub2D}, where $N$ denotes the number of OFDM symbols within the channel’s coherence time. 
Following \cite{WvH12}, the resulting \(M\times N\) time–frequency block is depicted in Fig.~\ref{fig:2dsym}.

\subsection{Gain and Variance Estimation}
Using the 2D block defined in Sec. ~\ref{sec:sub-2d}, we implement a channel gain estimator based on 2D blocks, which relies on the received signal power, similar to the method described in \cite{RRC03, WvH12} as
\begin{align}\label{eq:gain_est}
    \hat{G} &= \mathbb{E}\left\{ |Y_{m,n}|^2 \right\} - \hat{\sigma}_w^2,
\end{align}
where $m=1,2,\cdots,M$ and $n=0,1,\cdots,N-1$ are the indices for the 2D estimation block.
The variance in \eqref{eq:gain_est} is estimated blindly from the null tones in Fig.~\ref{fig:2DX}. Let $\mathcal{Y}_\varnothing$ denote the set of received null tones within one transmission frame. 
The noise variance is then estimated as
\begin{align}
    \hat{\sigma}^2_w 
    = \frac{1}{|\mathcal{Y}_\varnothing|} \sum_{Y_{m,n} \in \mathcal{Y}_\varnothing} |Y_{m,n}|^2.
\end{align}

\section{Numerical Simulations}\label{sec:results}
All simulation results are obtained from $1000$ independent Monte Carlo runs. A rate~$1/2$ convolutional code with generator polynomials $(133,171)_8$ is applied. A random bit-interleaver of the same length as the codeword is used, and the interleaved bits are subsequently mapped to QPSK symbols, which are then processed by a differential encoder. We consider the DAB Mode~I OFDM \cite{dab} format with a DFT size $M_f = 2048$. The active subcarriers $M_a = 1536$ follow the baseband index selection
\[
 \left\{k : \tfrac{M_f}{8}+1 \le k \le \tfrac{M_f}{2}\right\}
  \;\cup\;
  \left\{k : \tfrac{M_f}{2}+2 \le k \le \tfrac{7M_f}{8}+1\right\},
\]
which corresponds to nulled DC and left/right guard bands. The maximum number of OFDM symbols per transmission frame is 
$N_s = 19$, where each symbol is preceded by a cyclic prefix of length $N_{cp} = 504$. Time interleaving is applied across 16 consecutive transmission frames, which results in an interleaver depth (total codeword length) $N_T = 16 \cdot (N_s - 1) M_a \log_2(Q) = 16 \times 55296$.
The subcarrier spacing is $\Delta f = 1~\text{kHz}$, which yields the useful symbol duration $T_u = 1/\Delta f = 1~\text{ms}$. The cyclic prefix corresponds to a guard duration $T_{cp} =  T_u\cdot{N_{cp}}/{M_f} \approx 0.246~\text{ms}$, so that each OFDM symbol has a total duration $T_s = T_u + T_{cp} \approx 1.246~\text{ms}$. The sampling rate is $f_s = {M_f}/{T_u} = 2.048~\text{MHz}$.  Active subcarriers occupy the bandwidth $BW = M_a \Delta f = 1536 \cdot 1~\text{kHz} = 1.536~\text{MHz}$.

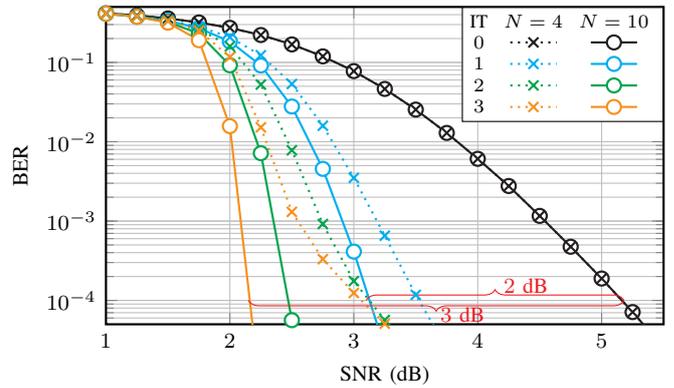
\begin{figure}[t]
    \centering
    \resizebox{1\columnwidth}{!}{\begin{tikzpicture}
	
	\begin{semilogyaxis}[
		axis line style = thick,
		grid = both,
		name = p1,
		width=1\columnwidth,
		height=.65\columnwidth,
		xmin = 1, xmax=5.5,
		ymin = 5e-5, ymax=0.5,
		font=\footnotesize,
		ylabel = BER, xlabel = SNR (dB),
		legend style={
			font=\tiny,
			nodes={scale=1.0},
		},
		legend cell align={left},
		legend pos = south west,
		]
        \addplot[color=Black, thick, mark=*, mark options={solid,fill=white}, mark size=2.6] 
             table[x expr=\thisrowno{0}, y expr=\thisrowno{1}] 
            {data/ber_N10_K7.txt};\label{bper0} 
		
        \addplot[color=Cyan, thick,mark=*, mark options={solid,fill=white}, mark size=2.6] 
             table[x expr=\thisrowno{0}, y expr=\thisrowno{2}] 
            {data/ber_N10_K7.txt};\label{bper1} 

        \addplot[color=Green, thick,mark=*, mark options={solid,fill=white}, mark size=2.6] 
             table[x expr=\thisrowno{0}, y expr=\thisrowno{3}] 
            {data/ber_N10_K7.txt};\label{bper2} 

        \addplot[color=BurntOrange, thick,mark=*, mark options={solid,fill=white}, mark size=2.6] 
             table[x expr=\thisrowno{0}, y expr=\thisrowno{4}] 
            {data/ber_N10_K7.txt};\label{bper3} 
            
        \addplot[color=Black, dotted, thick, mark=x, mark options={solid}, mark size=2.6] 
             table[x expr=\thisrowno{0}, y expr=\thisrowno{1}] 
            {data/ber_N4_K7.txt};\label{unc0} 
		
        \addplot[color=Cyan, dotted, thick, mark=x, mark options={solid}, mark size=2.6] 
             table[x expr=\thisrowno{0}, y expr=\thisrowno{2}] 
            {data/ber_N4_K7.txt};\label{unc1} 

        \addplot[color=Green, dotted, thick, mark=x, mark options={solid}, mark size=2.6] 
             table[x expr=\thisrowno{0}, y expr=\thisrowno{3}] 
            {data/ber_N4_K7.txt};\label{unc2} 

        \addplot[color=BurntOrange, dotted, thick, mark=x, mark options={solid}, mark size=2.6] 
             table[x expr=\thisrowno{0}, y expr=\thisrowno{4}] 
            {data/ber_N4_K7.txt};\label{unc3} 
            
        \draw[decorate,decoration={brace,amplitude=4pt},Red] (axis cs:3.1, 1e-4) -- (axis cs:5.18, 1e-4) node[pos=0.62, above=-1pt] {$2$~dB};
        \draw[decorate,decoration={brace,amplitude=4pt,mirror},Red] (axis cs:2.15, 1e-4) -- (axis cs:5.18, 1e-4)node[pos=0.57, below=-1pt] {$3$~dB};
                

	\end{semilogyaxis}
 
\matrix[
    matrix of nodes,
    anchor=north east,
    fill = white, draw,
    inner sep = 0.2em,
    column sep = 0.1em,
    node font=\scriptsize,
    column 1/.style={nodes={align=center}},
    column 2/.style={nodes={align=center}},
    column 3/.style={nodes={align=center}},
    column 4/.style={nodes={align=center}}
  ]
  at (current axis.north east){
    IT  & $N=4$   & $N=10$ \\ 
    $0$ & \ref{unc0} & \ref{bper0} \\
    $1$ & \ref{unc1} & \ref{bper1} \\
    $2$ & \ref{unc2} & \ref{bper2} \\
    $3$ & \ref{unc3} & \ref{bper3} \\};

\end{tikzpicture}    }
    \caption{BER performance over iterations (IT) of the optimal receiver design for inner code length $N=4$ (dotted lines) and $N=10$ (solid lines) over the AWGN channel.}   
    \label{fig:ber_awgn}
\end{figure}

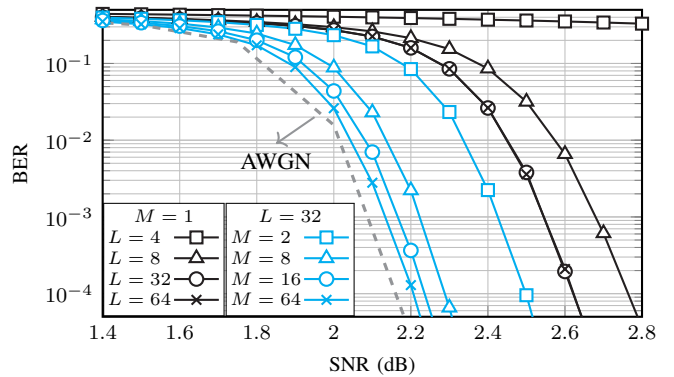
\begin{figure}[t]
    \centering
    \resizebox{1\columnwidth}{!}{\begin{tikzpicture}
	
	\begin{semilogyaxis}[
		axis line style = thick,
		grid = both,
		name = p1,
		width=1\columnwidth,
		height=.65\columnwidth,
		xmin = 1.4, xmax=2.8,
		ymin = 5e-5, ymax=0.5,
		font=\footnotesize,
		ylabel = BER, xlabel = SNR (dB),
		legend style={
			font=\tiny,
			nodes={scale=1.0},
		},
		legend cell align={left},
		legend pos = south west,
		]
        \addplot[color=Black, thick, mark=square*, mark options={solid,fill=white}, mark size=2.3] 
             table[x expr=\thisrowno{0}, y expr=\thisrowno{1}] 
            {data/ber_Np_M_K7.txt};\label{np0} 
		
        \addplot[color=Black, thick,mark=triangle*, mark options={solid,fill=white}, mark size=3] 
             table[x expr=\thisrowno{0}, y expr=\thisrowno{2}] 
            {data/ber_Np_M_K7.txt};\label{np1} 

        \addplot[color=Black, thick, mark=*, mark options={solid,fill=white}, mark size=2.6] 
             table[x expr=\thisrowno{0}, y expr=\thisrowno{3}] 
            {data/ber_Np_M_K7.txt};\label{np2} 

        \addplot[color=Black, thick,mark=x, mark options={solid,fill=white}, mark size=2.6] 
             table[x expr=\thisrowno{0}, y expr=\thisrowno{4}] 
            {data/ber_Np_M_K7.txt};\label{np3} 
            
        \addplot[color=Cyan, thick, mark=square*, mark options={solid,fill=white}, mark size=2.3] 
             table[x expr=\thisrowno{0}, y expr=\thisrowno{5}] 
            {data/ber_Np_M_K7.txt};\label{mm0} 
		
        \addplot[color=Cyan, thick, mark=triangle*, mark options={solid,fill=white}, mark size=3] 
             table[x expr=\thisrowno{0}, y expr=\thisrowno{6}] 
            {data/ber_Np_M_K7.txt};\label{mm1} 

        \addplot[color=Cyan, thick, mark=*, mark options={solid,fill=white}, mark size=2.6] 
             table[x expr=\thisrowno{0}, y expr=\thisrowno{7}] 
            {data/ber_Np_M_K7.txt};\label{mm2} 

        \addplot[color=Cyan, thick, mark=x, mark options={solid,fill=white}, mark size=2.6] 
             table[x expr=\thisrowno{0}, y expr=\thisrowno{8}] 
            {data/ber_Np_M_K7.txt};\label{mm3} 

        \addplot[color=Gray, dashed, very thick] 
             table[x expr=\thisrowno{0}, y expr=\thisrowno{4}] 
            {data/ber_N10_K7.txt};\label{bawgn} 
        \draw[->, Gray, thick] 
            (axis cs: 1.95, 2*10^-2) 
            -- (axis cs: 1.85, 9*10^-3); 
        \node at (axis cs: 1.85, 9*10^-3) [below] {\small{AWGN}};

	\end{semilogyaxis}

\node[anchor=south west, inner sep=0pt] at (current axis.south west) (m1) {%
  \scriptsize
  \setlength{\fboxsep}{0.2em}
  \setlength{\fboxrule}{0.4pt}
  \fcolorbox{black}{white}{%
    \begin{tabular}{@{}l@{\hspace{0.1em}}l@{}}
      \multicolumn{2}{c}{$M=1$}\\
    $L=4$  & \ref{np0}   \\
    $L=8$  & \ref{np1}  \\
    $L=32$ & \ref{np2}\\
    $L=64$ & \ref{np3}  \\
    \end{tabular}%
  }%
};
\node[anchor=west, inner sep=0pt] at (m1.east) (m2) {%
  \scriptsize
  \setlength{\fboxsep}{0.2em}
  \setlength{\fboxrule}{0.4pt}
  \fcolorbox{black}{white}{%
    \begin{tabular}{@{}l@{\hspace{0.1em}}l@{}}
      \multicolumn{2}{c}{$L=32$}\\
    $M=2$  & \ref{mm0}  \\
    $M=8$  & \ref{mm1} \\
    $M=16$ & \ref{mm2} \\
    $M=64$ & \ref{mm3} \\
    \end{tabular}%
  }%
};
 

    
\end{tikzpicture}    }
    \caption{BER performance of the joint phase estimation and demodulation design for different phase quantization levels (black solid lines) and subcarrier length $M$ (cyan solid lines) over the AWGN channel with a block-wise constant channel phase with inner code length $N=10$ and $3$ iterations.}   
    \label{fig:ber_phs}
\end{figure}
We begin by evaluating the optimal iterative DE-PSK receiver described in Sec.~\ref{sec:opt_rec} over the AWGN channel. For all simulations, we define the $\text{SNR}= E\{|X_t|^2\}/\sigma_w^2$. In Fig.~\ref{fig:ber_awgn}, the impact of inner code length (i.e., DE-QPSK symbol length $N$) is shown to significantly influence the performance improvement achieved through iterative demodulation and decoding. For $N = 4$, the iterative gain over $3$ iterations is approximately $2$~dB to reach a bit error rate (BER) of $10^{-4}$, while for $N = 10$, the gain increases to nearly $3$~dB. Fig.~\ref{fig:ber_phs} illustrates the estimation performance of the 2D blind receiver under varying phase quantization levels and subcarrier lengths in an AWGN environment with a block-invariant channel phase. Quantization with $L=8$ levels already provides a marked improvement over the baseline trellis with $L=Q=4$, while additional gains saturate for $L \geq 32$. Increasing the subcarrier length $M$ further improves the phase estimation performance. We show in Fig.~\ref{fig:ber_phs} that the blind receiver approaches optimal performance for $M = 64$. 

\begin{figure}[t]
    \centering
    \resizebox{1\columnwidth}{!}{\begin{tikzpicture}
	
	\begin{semilogyaxis}[
		axis line style = thick,
		grid = both,
		name = p1,
		width=1\columnwidth,
		height=.65\columnwidth,
		xmin = 2, xmax=8,
		ymin = 5e-5, ymax=0.5,
		font=\footnotesize,
		ylabel = BER, xlabel = SNR (dB),
		legend style={
			font=\tiny,
			nodes={scale=1.0},
		},
		legend cell align={left},
		legend pos = south west,
		]
        \addplot[color=Cyan, thick, mark=*, mark options={solid,fill=white}, mark size=2.3] 
             table[x expr=\thisrowno{0}, y expr=\thisrowno{1}] 
            {data/ber_tu6_ideal_i3_K7.txt};\label{per1}

        \addplot[color=Cyan, thick, mark=square*, mark options={solid,fill=white}, mark size=2.1] 
             table[x expr=\thisrowno{0}, y expr=\thisrowno{2}] 
            {data/ber_tu6_ideal_i3_K7.txt};\label{per2}

        \addplot[color=Cyan, thick, mark=triangle*, mark options={solid,fill=white}, mark size=3] 
             table[x expr=\thisrowno{0}, y expr=\thisrowno{3}] 
            {data/ber_tu6_ideal_i3_K7.txt};\label{per3}
            
        \addplot[color=Cyan, thick, dotted, mark=*, mark options={solid,fill=white}, mark size=2.3] 
             table[x expr=\thisrowno{0}, y expr=\thisrowno{1}] 
            {data/ber_tu6_ideal_i0_K7.txt};
        \addplot[color=Cyan, thick,dotted, mark=square*, mark options={solid,fill=white}, mark size=2.1] 
             table[x expr=\thisrowno{0}, y expr=\thisrowno{2}] 
            {data/ber_tu6_ideal_i0_K7.txt};
        \addplot[color=Cyan, thick,dotted, mark=triangle*, mark options={solid,fill=white}, mark size=3] 
             table[x expr=\thisrowno{0}, y expr=\thisrowno{3}] 
            {data/ber_tu6_ideal_i0_K7.txt};
            
        \addplot[color=Black, thick, mark=*, mark options={solid,fill=white}, mark size=2.3] 
             table[x expr=\thisrowno{0}, y expr=\thisrowno{1}] 
            {data/ber_tu6_i3_K7.txt};\label{bli1}

        \addplot[color=Black, thick, mark=square*, mark options={solid,fill=white}, mark size=2.1] 
             table[x expr=\thisrowno{0}, y expr=\thisrowno{2}] 
            {data/ber_tu6_i3_K7.txt};\label{bli2}

        \addplot[color=Black, thick, mark=triangle*, mark options={solid,fill=white}, mark size=3] 
             table[x expr=\thisrowno{0}, y expr=\thisrowno{3}] 
            {data/ber_tu6_i3_K7.txt};\label{bli3_K7}
            
        \addplot[color=Black, thick, mark=+, mark options={solid,fill=white}, mark size=2.3] 
             table[x expr=\thisrowno{0}, y expr=\thisrowno{4}] 
            {data/ber_tu6_i3_K7.txt};\label{bli4}
            
        \addplot[color=Black, thick, mark=asterisk, mark options={solid,fill=white}, mark size=2.3] 
             table[x expr=\thisrowno{0}, y expr=\thisrowno{5}] 
            {data/ber_tu6_i3_K7.txt};\label{bli5}

        \addplot[color=Black, thick, dotted, mark=*, mark options={solid,fill=white}, mark size=2.3] 
             table[x expr=\thisrowno{0}, y expr=\thisrowno{1}] 
            {data/ber_tu6_i0_K7.txt};

        \addplot[color=Black, thick, dotted, mark=square*, mark options={solid,fill=white}, mark size=2.1] 
             table[x expr=\thisrowno{0}, y expr=\thisrowno{2}] 
            {data/ber_tu6_i0_K7.txt};

        \addplot[color=Black, thick, dotted, mark=triangle*, mark options={solid,fill=white}, mark size=3] 
             table[x expr=\thisrowno{0}, y expr=\thisrowno{3}] 
            {data/ber_tu6_i0_K7.txt};
            
        \addplot[color=Black, thick, dotted, mark=+, mark options={solid,fill=white}, mark size=2.3] 
             table[x expr=\thisrowno{0}, y expr=\thisrowno{4}] 
            {data/ber_tu6_i0_K7.txt};
        \addplot[color=Black, thick, dotted, mark=asterisk, mark options={solid,fill=white}, mark size=2.3] 
             table[x expr=\thisrowno{0}, y expr=\thisrowno{5}] 
            {data/ber_tu6_i0_K7.txt};

        \coordinate (it0center) at (axis cs:7.3, 2.5e-3); 
        \draw [dashed, thick, Red, rotate=50] (it0center) ellipse ({0.7cm} and {0.3cm});
        \node[anchor=west, Red, fill=white, inner sep=1pt] at ([xshift=5pt, yshift=23pt]it0center) {IT0};

        \coordinate (it3center) at (axis cs:4.5, 1.1e-2); 
        \draw [dashed, thick, Red] (it3center) ellipse ({1.5cm} and {0.15cm});
        \node[anchor=west, Red, fill=white, inner sep=1pt] at ([xshift=-48pt, yshift=7pt]it3center) {IT3};
        
            
	\end{semilogyaxis}

\node[anchor=south west, inner sep=0pt] at (current axis.south west) (m1) {%
  \scriptsize
  \setlength{\fboxsep}{0.2em}
  \setlength{\fboxrule}{0.4pt}
  \fcolorbox{black}{white}{%
    \begin{tabular}{@{}l@{\hspace{0.1em}}l@{}}
    \multicolumn{2}{c}{\textbf{Blind}}\\
    N4\,\,\, M1 & \ref{bli1}  \\
    N7\,\,\, M1 & \ref{bli2}  \\
    N19 M1 & \ref{bli3_K7}  \\
    N7\,\,\, M8 & \ref{bli4}  \\
    N7\,\,\, M128 & \ref{bli5}  \\
    \end{tabular}%
  }%
};

\node[anchor=north east, inner sep=0pt] at (current axis.north east) {%
  \scriptsize
  \setlength{\fboxsep}{0.2em}
  \setlength{\fboxrule}{0.4pt}
  \fcolorbox{black}{white}{%
    \begin{tabular}{@{}l@{\hspace{0.1em}}l@{}}
      \multicolumn{2}{c}{\textbf{Ideal}}\\
      $N4$  & \ref{per1} \\
      $N7$  & \ref{per2} \\
      $N19$ & \ref{per3} \\
    \end{tabular}%
  }%
};
\end{tikzpicture}    }
    \caption{BER performance of the blind receiver design over the time-varying TU6 channel with different 2D $M\times N$ block size after $3$ iterations. Dotted lines indicate zero iteration, while solid lines represent the results after $3$ iterations.}   
    \label{fig:ber_tu6}
\end{figure}
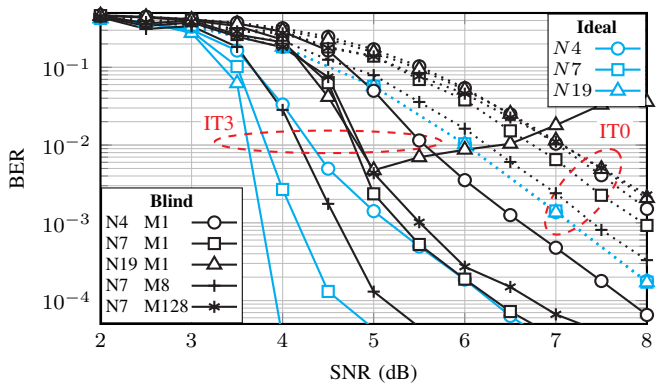

To assess the full capability of the blind receiver, we consider the typical urban $6$ taps (TU6) channel model, commonly used to evaluate the DAB transmission. In Fig.~\ref{fig:ber_tu6}, we demonstrate the performance of our fully blind turbo-DE-PSK receiver for a Doppler frequency of $10$~Hz, which corresponds to a speed of $45$~km/h for DAB. As a benchmark, the optimal receiver performance with full access to the channel knowledge is also presented (light blue curves). First, we evaluate the impact of the inner code length $N$. As expected, increasing the length from $N=4$ to $N=7$ improves the performance of the blind receiver. 
Nevertheless, when the inner code length exceeds the channel coherence time, the fixed channel estimates become systematically biased. At high SNR, this bias dominates and the iterative process reinforces incorrect likelihoods, causing the receiver to diverge instead of improving. On the other hand, increasing the number of subcarriers $M$ improves channel estimation and results in a better BER. However, when the number of subcarriers exceeds the channel coherence bandwidth, the performance of the blind receiver begins to degrade.

\section{Conclusions}\label{sec:conc}
We generalized and discussed the use-case scenarios and robustness of a fully blind turbo demodulation framework for DE OFDM systems that estimates channel phase, gain, and variance directly from the received signal without pilots. The receiver employs 2D trellis decomposition for joint phase estimation and symbol demodulation, while a power-based method provides blind gain and variance estimation for the likelihood function. Simulations with DAB-like formats showed that the blind receiver closely approaches the performance of receivers with perfect channel knowledge. The influence of inner code length, phase quantization, and 2D block size was also quantified.  

Future work will focus on enhancing receiver capabilities to handle highly time-varying channels and non-Gaussian noise environments. Overall, the results indicate that blind turbo demodulation is a promising approach for next-generation pilot-free OFDM broadcast systems.

\section*{Acknowledgments}
This work was funded by the RAISE collaboration framework between Eindhoven University of Technology and NXP, including a PPS-supplement from the Dutch Ministry of Economic Affairs and Climate Policy.

\vspace{12pt}
\end{document}